\documentclass[apjl]{emulateapj}

\newcommand{\ie}{i.e.,\ }
\newcommand{\eg}{e.g.,\ }

\newcommand{\etal}{et~al.\ }
\newcommand{\ltsima}{$\; \buildrel < \over \sim \;$}
\newcommand{\simlt}{\lower.5ex\hbox{\ltsima}}
\newcommand{\gtsima}{$\; \buildrel > \over \sim \;$}
\newcommand{\simgt}{\lower.5ex\hbox{\gtsima}}

\newcommand{\magsec}{mag/arcsec$^2$}
\newcommand{\alphadef}{$=N_{\rm PNe}/L_{\rm bol}$}

\def\parcmin{{\tt '}\mskip -6.0mu.\,}

\def\muv{$\mu_{\mbox{V}}$}
\def\mub{$\mu_{\mbox{B}}$}
\def\ncand{$N_{\rm c}$\ }

\begin{document}

\title{The Connection Between Diffuse Light and Intracluster Planetary Nebulae in
  the Virgo Cluster}

\author{J. Christopher Mihos\altaffilmark{1}, Steven Janowiecki\altaffilmark{1,2},
John J. Feldmeier\altaffilmark{3}, Paul Harding\altaffilmark{1}, and Heather Morrison\altaffilmark{1}}
\email{mihos@case.edu, steven.janowiecki@case.edu, jjfeldmeier@ysu.edu,
harding@case.edu, hlm5@case.edu}

\altaffiltext{1}{Department of Astronomy, Case Western Reserve University,
10900 Euclid Ave, Cleveland, OH 44106}

\altaffiltext{2}{now at the Department of Astronomy, Indiana
  University}

\altaffiltext{3}{Department of Physics and Astronomy, Youngstown State University,
Youngstown, OH 44555}

\begin{abstract}

We compare the distribution of diffuse intracluster light detected in
the Virgo Cluster via broadband imaging with that inferred from
searches for intracluster planetary nebulae (IPNe). We find a rough
correspondence on large scales ($\sim$ 100 kpc) between the two, but
with very large scatter ($\sim$ 1.3 mag/arcsec$^2$). On smaller scales (1
-- 10 kpc), the presence or absence of correlation is clearly
dependent on the underlying surface brightness. On these scales, we
find a correlation in regions of higher surface brightness (\muv
\simlt 27) which are dominated by the halos of large galaxies such as
M87, M86, and M84. In those cases, we are likely tracing PNe associated with
galaxies rather than true IPNe. In true intracluster fields, at lower
surface brightness, the correlation between luminosity and IPN
candidates is much weaker. While a correlation between broadband light
and IPNe is expected based on stellar populations, a variety of
statistical, physical, and methodological effects can act to wash out
this correlation and explain the lack of a strong correlation at lower
surface brightness found here. A significant complication comes from
the stochastic nature of the PN population combined with contamination
of the IPN catalogs due to photometric errors and background emission
line objects. If we attribute the lack of a strong observed
correlation solely to the effects of contamination, our Monte Carlo
analysis shows that our results are mostly consistent with a
``IPNe-follow-light'' model if the IPN catalogs are contaminated by
$\sim$ 40\%. Further complications arise from the line-of-sight depth
of Virgo and uncertainty in the stellar populations of the ICL, both
of which may contribute to the slight systematic differences seen
between the IPN-inferred surface brightnesses and those derived from
our deep surface photometry.

\end{abstract}


\section{Introduction}

The diffuse intracluster light (ICL) that permeates clusters of
galaxies has become a valuable tool for studying galaxy cluster
evolution. This ICL arises from a variety of sources, and is likely
dominated by tidal stripping of galaxies during interactions both
within the cluster and within infalling groups (Merritt 1984; Moore
\etal 1996; Rudick, Mihos, \& McBride 2006; Murante \etal 2008), as
well as stripping by the cluster potential itself for galaxies passing
near the cluster core (Byrd \& Valtonen 1990; Gnedin 2003). Recent
models of galaxy cluster accretion and evolution show that $\sim
10-40$\% of the luminosity of clusters may be found in the ICL (\eg
Willman \etal 2004; Murante \etal 2004; Sommer-Larsen \etal 2005;
Rudick \etal 2006; Purcell \etal 2007). Deep imaging of galaxy
clusters has begun to reveal this ICL component (Bernstein \etal 1995;
Feldmeier \etal 2002, 2004; Mihos \etal 2005; Zibetti \etal 2005;
Gonzalez \etal 2005), with some suggestion of a
correlation between cluster mass and ICL fraction (Ciardullo \etal
2004; Krick \& Bernstein 2007).

Of particular interest is the dynamical information contained within
the ICL. As material is stripped from its host galaxies, long tidal
streams are formed (Gregg \& West 1998; Calc\'aneo-Rold\'an \etal
2000; Feldmeier \etal 2004; Mihos \etal 2005); these coherent streams
can later be destroyed during a subsequent cluster accretion event,
contributing to a smooth diffuse ICL component (\eg Rudick \etal 2006,
2008). The morphology of the diffuse ICL on cluster scales thus
contains information on the recent accretion history of galaxy
clusters. On smaller scales, the ICL can trace the dynamical history
of individual galaxies, as extended tidal debris can highlight past
interactions, either with the cluster potential or with other galaxies
(\eg Schweizer 1980; Malin 1994; Weil \etal 1997; Katsiyannis \etal
1998; Feldmeier \etal 2004; Janowiecki \etal in preparation). 

Along with diffuse light, the ICL can be studied through discrete
tracers such as globular clusters (West \etal 1995; Williams \etal
2007a), red giants (Ferguson, Tanvir, \& von Hippel 1998; Durrell
\etal 2002; Williams \etal 2007b), and planetary nebulae (Arnaboldi
\etal 1996; Feldmeier \etal 1998, 2004; Aguerri \etal
2005). Intracluster planetary nebulae (IPNe) are of particular
interest here, since they are believed to trace the stellar luminosity
of galaxies and -- as emission line objects -- they permit study of
the kinematics of the ICL. IPN kinematics have been studied in both
the Virgo (Arnaboldi \etal 2004) and Coma (Gerhard \etal 2005)
clusters, and yield interesting constraints on kinematic
substructure. Larger samples of IPN velocities would yield information
on the degree of relaxation of the ICL, its kinematic connection to
the cluster galaxies, and also provide further constraints on the
cluster mass distribution (Gerhard \etal 2007; Sommer-Larsen \etal
2005; Willman \etal 2005).  However, existing surveys for IPNe have
largely been done ``blind,'' without any of the foreknowledge of the
underlying properties of the diffuse light provided by recent
broadband imaging surveys (\eg Mihos \etal 2005). A strong correlation
between broadband light and IPN density would then open up the
capability of using deep imaging to maximize the efficiency of
followup IPN searches.

If well determined, the connection between planetary nebulae and
broadband luminosity in the ICL also can provide constraints on its
stellar populations. The ratio $\alpha=N_{\rm PNe}/L_{\rm bol}$
depends on the stellar populations of the ICL (see, \eg Buzzoni,
Arnaboldi, \& Corradi 2006), and in galaxies shows a correlation with
galaxy type (Peimbert 1990; Ciardullo \etal 2005), wherein massive
metal-rich ellipticals have a much lower value of $\alpha$ than do
low-luminosity, metal-poor galaxies. Measurements of $\alpha$ can test
various models for the formation of the ICL, as ICL formed from the
mergers of massive galaxies during the formation of a cluster (Murante
\etal 2008) would show lower values of $\alpha$ than ICL formed from
the ongoing disruption of low luminosity dwarfs orbiting in the
cluster potential (Moore \etal 1996).

The data now exist to do this detailed comparison between the ICL
measured by broad band light and IPNe in the Virgo Cluster. We (Mihos
\etal 2005) have conducted a deep wide-field imaging survey of the
Virgo core down to a $3\sigma$ limiting surface brightness of $\mu_V =
28.5$, while narrowband surveys (Feldmeier \etal 2004; hereafter F04;
Aguerri \etal 2005; hereafter A05) are discovering significant numbers
of IPN candidates in Virgo. Many of these IPN fields overlap with
our imaging fields, allowing us to do direct comparison between the
different ICL tracers. These tracers have very different systematic
uncertainties: broad band surface photometry is sensitive to sky
subtraction, flat field quality, and scattered light control, while
the IPN measurements rely on statistics of faint point source
detection and contamination from background emission line
objects. While the comparison will be complicated by these different
systematic uncertainties, it also gives us a new opportunity to assess
the consistency of the two methods in detecting the diffuse
intracluster light in galaxy clusters.

\section{Observational Data}

\subsection{Broad band imaging}

\begin{figure*} \figurenum{1}
\plotone{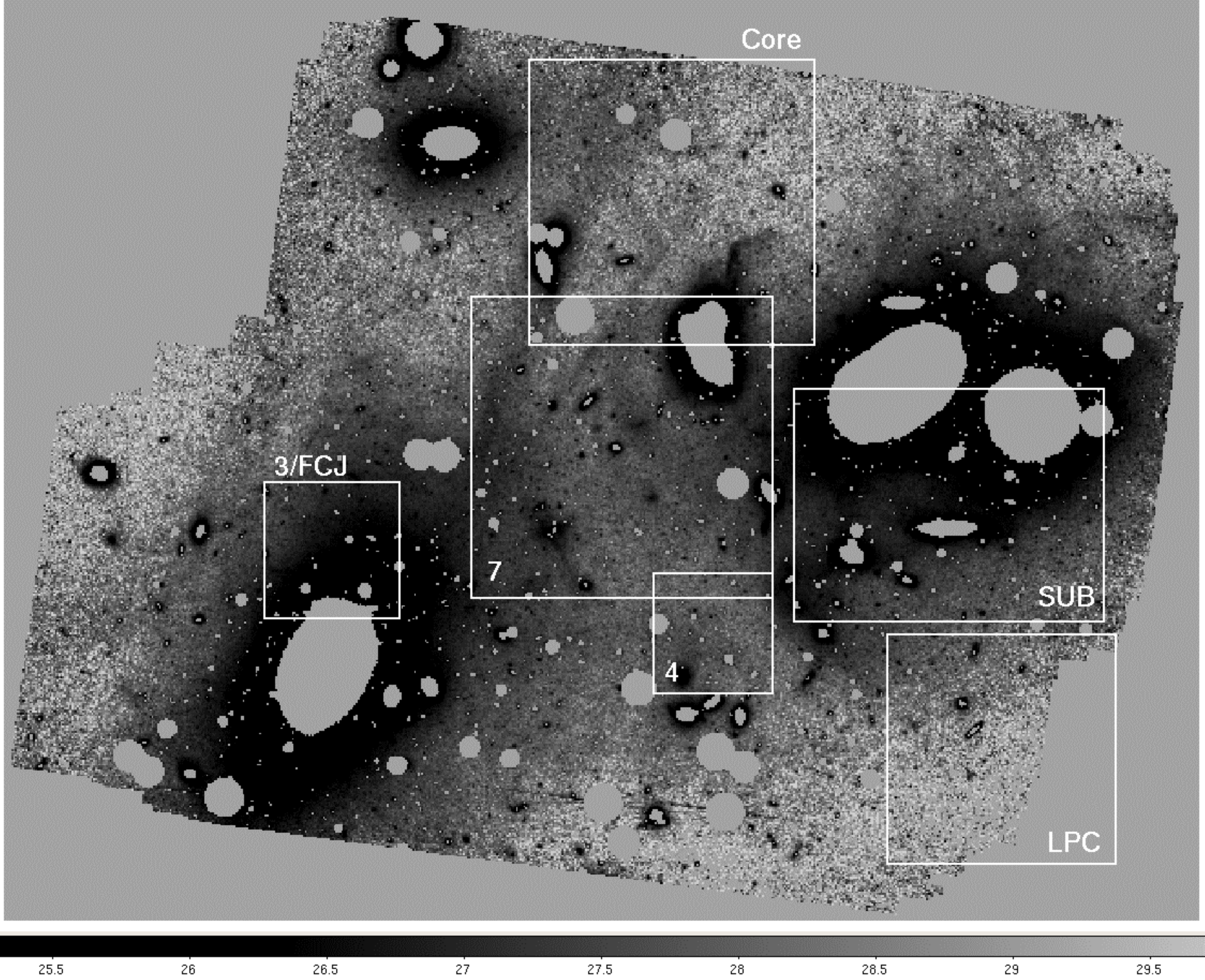}
\caption{
Optical image of the Virgo cluster core, taken from Mihos \etal
2005. Intensity saturates black at a surface brightness of \muv=26.5,
and the faintest ICL features visible have surface brightness \muv
$\sim$ 28.5. The white masked areas represent either foreground
stars or the inner regions of galaxies brighter than \muv=25. The
various IPN fields from F04 (fields 3, 4, and 7) and A05 (FCJ, Core,
SUB, and LPC) are overlaid.
\label{fig:fields}}
\end{figure*}

The optical imaging comes from our ongoing survey for ICL in the Virgo
cluster using the 0.6m Burrell Schmidt at Kitt Peak. Figure 1 shows
our image of the central 2.5 square degrees of the Virgo core (see
Mihos \etal 2005 for details). This image has been constructed from
co-adding 72 individual 900s images, each dithered by up to half a
degree. The dithering pattern means that the total exposure time
across the field varies. The median number of exposures per pixel in
the final mosaic is 31. The exposure time is highest in the inner 17\%
of the image, where there are at least 40 exposures per pixel; 73\% of
the pixels have at least 20 exposures, and the minimum exposure per
pixel (which defines the edges of the mosaic) is set at 5. Each
individual image was flattened using a night sky flat built from 127
offset sky pointings. The sky pointings were taken bracketing the
Virgo images in both time and position, to minimize flat fielding
variations due to telescope flexure and illumination. All images were
taken in Washington M and transformed to Johnson V assuming a color of
B-V=1 typical of the old stellar populations thought to represent
intracluster light. The uncertainty in the ICL surface brightness
introduced by the photometric transform is small -- for example,
adopting an extremely and unrealistically blue ICL color of B-V=0
changes the inferred surface brightness by only 0.18 magnitudes.

Sky subtraction is problematic when dealing with an image which
subtends only a small fraction of the cluster area. To obtain an
reliable estimate of sky, we first need to subtract the wings of
bright stars from the image. We use long exposures of $\alpha$ Leo to
construct the stellar point spread function out to 30\arcmin, and then
subtract the extended wings of stars out a radius where they
contribute 0.1 ADU ($\mu_V=30$) to the local sky. We also mask stars
completely inside the radius where the scaled stellar PSF is brighter
than 3 ADU.  IRAF's objmask task is then run on the star subtracted
image to mask all {\it extended} sources 2.5$\sigma$ above a
locally-measured sky value. The image is then spatially rebinned into
``superpixels'' which are 32x32 pixels in size; each superpixel is
given an intensity equal to the mode of its individual pixels. Regions
are then selected to represent ``true sky'' -- these regions are
chosen to be far from bright galaxies, and typically make up $\sim$
15\% of each individual image. For each image, an absolute sky value
is then calculated from the mode of the superpixel intensities in
these pure sky regions and subtracted off. Depending on varying
airmass, time, and hour angle, these sky levels varied from 1100-1500
ADU ($\mu_V = 21.6 - 21.3$).

To make the final mosaic, we then register each image and use the
``true sky'' regions to simultaneously and iteratively fit residual
sky planes to the individual images, constrained to minimize the
frame-to-frame deviations in overlapping regions of images. After
registering and medianing the individual images into the final mosaic,
one additional residual sky plane is fit and subtracted from the
mosaic. It is important to note that this process will over-subtract
any ICL which is smoothly distributed on scales {\it larger} than the
areal coverage of the image, approximately 400 kpc on a
side\footnote{We adopt a Virgo distance of 16 Mpc throughout this
paper. At this distance, 1\arcmin = 4.65 kpc}.

To construct the error model for the dataset, we follow the procedure
outlined by Morrison \etal (1997), described in more detail in the
Appendix. In brief, we expect a random error of approximately 2 ADU
per pixel\footnote{The photometric solution is such that 1 ADU
corresponds to a surface brightness of $\mu_V = 29.1$ mag/arcsec$^2$.}
from a combination of readnoise, small-scale flat fielding errors, and
sky photon noise. This uncertainty should become negligible when
measuring the diffuse ICL over large scales ($>$30\arcsec), where many
pixels contribute to the measurement. On these larger scales, however,
the error model is dominated by uncertainty in the sky background and
large scale flat-fielding, which we estimate to be approximately 0.5
ADU by examining the residual variance in boxes $2\parcmin5$ on a side
in regions lacking galaxies or diffuse light. This places the
$3\sigma$ limiting depth to our surface photometry at $\mu_V=28.65$.

\begin{deluxetable*}{cccccccccc}
\tabletypesize{\scriptsize}
\tablecaption{Virgo IPNe fields}
\tablewidth{0pt}
\tablehead{
\colhead{Field} & \colhead{$m_{lim}$} & \colhead {$d_{lim}$} & 
\colhead{$N_{cand}$} & \colhead{\muv} & \colhead{Source} &
\colhead{} &  \colhead{\muv (phot)} & \colhead{} & \colhead{$N_{images}$}\\
\colhead{} & \colhead{(5007)} & \colhead{(Mpc)} &  \colhead{$<m_{lim}$} &
\colhead{(PNe)} & \colhead{} & \colhead{$\mu_{\rm mask}$} &
\colhead{$\mu_{\rm mask}$} & \colhead{$\mu_{\rm mask}$} & \colhead{} \\
\colhead{} & \colhead{} & \colhead{(a)} & \colhead{} &
\colhead{(b)} & \colhead{} & \colhead{25.5} &  \colhead{26} &
\colhead{26.5} & \colhead{(c)}
}

\startdata 
  3  & 27.0 & 20.0 & 46 & 26.5 & F04 & 26.4 & 26.7 & 27.0 & 35.3\\
  4  & 26.6 & 16.6 & 4 & 27.4 & F04 & 27.6 & 27.7 & 27.8 & 34.2\\
  7  & 26.8 & 18.3 & 32 & 28.4 & F04 & 27.2 & 27.3 & 27.5 & 45.1\\
Core & 27.2 & 22.0 & 77 & 28.6\tablenotemark{c} & A05 & 27.8 & 27.9 & 28.1 & 35.2\\
 FCJ & 27.0 & 20.0 & 20 & 27.4\tablenotemark{c} & A05 & 26.4 & 26.7 & 27.0 & 35.3\\
 LPC & 27.5 & 25.2 & 14 & 31.7\tablenotemark{c} & A05 & 28.1 & 28.2 & 28.4 & 16.6\\
 SUB & 28.1 & 33.1 & 36 & 28.5\tablenotemark{c} & A05 & 26.6 & 26.9 & 27.3 & 37.7\\
\enddata
\tablenotetext{}{(a): $d_{lim}$ is the distance out to which a PNe at the
  high luminosity cutoff of the PNLF ($M^*=-4.51$) could be detected given the limiting apparent
  magnitude of the field.}
\tablenotetext{}{(b): \muv\ for A05 fields were converted from their
  quoted \mub\ by adopting an ICL color of B-V=1.2}
\tablenotetext{}{(c): For each IPN field, the average number of 900s exposures 
  contributing to the deep imaging mosaic of Mihos \etal (2005).}
\end{deluxetable*}

\subsection{IPN catalogs}

We compare the broadband diffuse light measured in our Virgo imaging
to that inferred from compilations of recent Virgo IPN surveys by F04
and A05. Catalogs for the latter datasets were kindly provided by
Ortwin Gerhard and collaborators. These surveys were conducted on a
variety of telescopes, have different areal coverage and selection
effects, and go to different limiting magnitudes; combining these
catalogs into a homogeneous dataset is problematic. Instead, our goal
is to compare the IPN-inferred ICL surface brightnesses averaged over
each field (\S3), and then examine the correlation between individual
IPN candidates and broadband light within individual fields (\S4).

One major issue is that these catalogs are catalogs of IPN {\it candidates}
detected via a two filter method. IPN candidates are identified as
point sources that are present in a narrowband ``on-band'' image that
includes the redshifted $\lambda$ 5007 [OIII] line, but disappears in
a spectrally-adjacent ``off-band'' image that excludes the
line. Unfortunately, such a search technique can also include
contaminating objects, and the contamination in the IPN catalogs can
be significant.  This contamination comes in two forms. The first is
background emission line objects, most notably high-redshift galaxies.
At a redshift of $z=3.1$, Ly$\alpha$ falls into the narrowband [OIII]
filters used to detect IPNe, such that a fraction of the IPN
candidates are in fact background Ly$\alpha$ emitting
galaxies. Estimates of this contamination fraction from narrowband
imaging of blank fields (Ciardullo \etal 2002) put the fraction of
background emitters at $\sim$ 20\%. This fraction can vary
significantly from field to field, however, depending on the
fluctuations in the background distribution of the Ly$\alpha$
emitters.  A second source of contamination can occur if the off-band
image has a shallow limiting magnitude and joint photometric errors in
the on-band and off-band images lead to sources being erroneously
classified as emission-line objects (the so-called ``spillover
effect,'' see A05).  Combined, these effects can lead to significant
contamination of the catalogs: Arnaboldi \etal (2004) have performed
follow-up spectroscopy of IPN candidates in the FCJ, Core, and Sub
fields, and achieve a confirmation rate of 84\%, 32\%, and 72\%,
respectively, for IPN candidates brighter than $m_{5007}=27.2$. We
address the issue of contamination in more depth in \S4 and \S5.

A second concern is the uniqueness of the IPN candidate lists. Both
the Field 3 and Field FCJ candidates were extracted from the same
imaging dataset, but by two different groups using somewhat different
methods for extracting the emission line candidates (see F04 and A05
for details). An examination of the candidates in the two fields shows
markedly different samples. Of the 46 Field 3 and 19 FCJ candidates
brighter than $m_{lim, 5007}=27$, only 12 candidates are common to
both catalogs. Spectroscopic follow-up of the FCJ candidates has been
conducted by Arnaboldi \etal (2004), and the overlap of the
spectroscopically-confirmed candidates is much better: 10/11
spectroscopically confirmed FCJ candidates also show up in the Field 3
catalog. Nonetheless, without full spectroscopic follow-up of all the
IPN candidates in survey fields, we limit our analysis to the existing
photometric IPN candidate lists.

With these caveats in mind, we show in Figure 2 the IPN candidates
from F04 and A05 overlaid on the broadband image from Mihos \etal
(2005). A visual examination of Figure 2 suggests that the correlation
between PNe candidates and diffuse light depends on the underlying
surface brightness -- the connection is stronger near the luminous
Virgo ellipticals, where their halo light dominates the fields, and
weak or absent in regions of low surface brightness in truly
intracluster fields. For example Field 3/FCJ, which is dominated by
halo light from M87, shows a clear gradient in PNe density from the NE
to the SW, along the direction of M87's luminosity gradient. Many of
these PNe are likely associated with M87 itself rather than being true
intracluster PNe (see, \eg the discussion in F04); indeed, followup
spectroscopy of 15 IPN candidates in the FCJ field by Arnaboldi \etal
(2004) shows that 12 of the 15 candidates have a narrow velocity
distribution ($\sigma=247 \pm 52$ km/s) that can be associated with
M87's halo. Similarly, the majority of IPN candidates in Field SUB are
very clearly concentrated around the giant ellipticals M86 and M84.

In other fields further away from the luminous ellipticals and at much
lower surface brightness, the correspondence between IPN candidates
and broad band light is less clear.  Field 4 only has four IPN
candidates brighter than the limiting magnitude; it is hard to make
any inferences about this field with so few candidates. Field 7 is
more complex; there are candidate IPNe around the galaxy pair NGC
4435/8, as well as along the eastern edge of the field, where they may
be associated with tidal streams coming from M87's halo. In the Core
field, there are many IPN candidates but no obvious connection to the
galaxies.  Field LPC is located in a region of our image that shows
the least diffuse light, and indeed has only a small number of IPN
candidates. If anything, however, these candidates appear
anti-correlated with the broadband light, as they lie in the southwest
(lower right) portion of the field while the M86/M84 complex lies just
off the northeast (upper left) corner of the field.

As a comparison, we show in Figure 3 {\it simulated} PNe catalogs for
each survey field, based on the assumption that PNe follow the
broadband light detected in our deep imaging. To construct these
catalogs, we build for each field a master PNe list under the
assumption that the number of PNe in any given pixel is proportional
to the pixel intensity ($N_{PNe} = C\times I_{i,j}$). To avoid PNe
being assigned to noise spikes, the images are all median smoothed
with a 9x9 pixel box before the IPN catalog is generated, and we do
not generate PNe in regions of high surface brightness (at $\mu_V<25$
for all fields but SUB; in SUB, PNe candidates were found at higher
surface brightness, and we choose the cutoff for this field to be
$\mu_V<24$.) We choose the constant of proportionality $C$ so that a
master list is generated of $O(10^6)$ PNe, which ensures that master
catalog contains potential IPN candidates even at low surface
brightnesses. We then randomly select from this master catalog the
same number of IPNe as found in the actual narrowband imaging data
from F04 and A05. The random catalog shown in Figure 3 is typical of
the catalogs generated this way.

Examining Figure 3, the connection between the ICL and the simulated
IPNe is clearly stronger than in the real data. Many simulated PNe are
found near the bright ellipticals in Fields 3, FCJ, and SUB, but now
the correlation with NGC 4435/8 is much stronger in Fields 7 and
Core. The simulated IPNe in the eastern (left) half of Field 7 more
obviously lie on top of the M87 streams, and the simulated IPNe in
Field LPC are concentrated in the northeast (upper left) portion of
the image, towards the M86/M84 pair. The fact that these spatial
biases are evident in the simulated dataset shows that the weaker bias
seen in the real dataset is likely not simply a result of small number
statistics.

Visually, therefore, the correlation between the IPN candidates and
the diffuse broadband light is loose at best. In regions of higher
surface brightness which encompass the extended halos of bright
ellipticals -- Field 3/FCJ and Core -- the IPN candidates do seem to
correlate with the galaxy light. At lower surface brightness in
regions which are more truly intracluster, the results are more
ambiguous -- Field 7 suggests a possible correlation between the
diffuse light and IPN candidates, but this not seen in Fields Core and
LPC. Because of the lower surface brightness in the intracluster
fields, much of this lack of correlation may be due to fractional
contamination being higher in these fields. Furthermore, other factors
may contribute as well to wash out the correlation, such as the depth
of the Virgo Cluster causing IPNe on the far side of Virgo to be lost
from the magnitude-limited IPN samples, or systematic variations in
the value of $\alpha$ across Virgo. We address each of these
possibilities in more detail below.  However, because visual
impressions are easily misled by the eye's desire to find patterns in
the data, we first must develop a more quantitative measure of the
correlation between IPNe and the diffuse light.

Ideally, one could do this comparison by combining the various IPN
catalogs into a single master catalog, and calculating the surface
density of IPN candidates in regions of a given surface brightness
across the cluster as whole. Unfortunately, this test is problematic
for a number of reasons -- one being the heterogeneity of the existing
IPN datasets. Because the datasets all have different limiting
magnitudes, contamination fractions, and selection criteria, we cannot
combine them together to generate a homogeneous "master list" of IPN
candidates on which to make the suggested test. For example, at a
given surface brightness, the IPN counts in one field may be low
compared to another simply because the first field did not go as
deep. Alternatively, counts in one field could be high compared to
another if the first field had more contaminants.  In principle, while
a homogeneous master catalog cannot be built, we could attempt the
test within individual fields. Here, however, we are limited both by
the small number of IPN candidates per field, and, at lower surface
brightness, uncertainty in the absolute surface photometry due to sky
subtraction.

In Fields 3/FCJ and SUB, however, we have a means to attempt this test
in a somewhat more model-dependent fashion. These fields are dominated
by the halos of the elliptical galaxies M87 (for Field 3/FCJ) and
M84/M86 (for Field SUB), and using our deep imaging we can use IRAF's
{\tt ELLIPSE} task to construct model luminosity profiles for these
ellipticals (Janowiecki \etal in preparation). We then define the
areas of the field covered by isophotes of a given surface brightness,
and calculate the IPN surface density in these areas as a function of
surface brightness. Since the isophotal models are spatially smooth
and extend to high surface brightness, this should provide a
demonstration of how well the IPN candidates follow the galaxy
luminosity distribution. The results are presented in Figure 4, where
the dotted line shows the expectation if PNe surface density follows
the stellar light profile of the galaxies. The correlation between the
galaxy light and PNe surface density is clear, reminiscent of other
studies of the PNe distribution around luminous spheriods such as M104
(Ford \etal 1996) and NGC 5128 (Hui \etal 1993). 

However, the analysis in Figure 4 only tests the correlation between
IPN density and the light profile of the giant ellipticals. It does
not probe truly intracluster regions at low surface brightness where
the luminosity distribution is spatially complex, nor does it factor
in the contribution of fainter galaxies which may lie in the field. To
test the correlation between IPN candidates and diffuse light more
generally in all the IPN survey fields, we need to employ an
alternative set of quantitative tests to explore the correlation
between IPN candidates and broad band light. First, in \S3 we compare
the {\it absolute} level of surface brightnesses measured by our deep
imaging with that inferred from the IPNe surveys, averaged over each
survey field.  Then, in \S4 we
examine the {\it relative} excess of light around IPN candidates in
individual fields. The first approach deals with the problems of
heterogeneity in the ICL data by relying on the statistical
corrections for completeness and contamination that F04 and A05 have
adopted for their datasets, allowing us to make the proper absolute
comparison between different survey fields. The second approach deals
with the problems of systematic uncertainty in the surface brightness
data by making a relative measure of flux excess rather than an
absolute measurement of surface brightness. If IPNe faithfully trace
the diffuse light, that should manifest as an excess of light in
regions around IPN candidates, compared to randomly sampled regions
throughout a given field.

\begin{figure*} \figurenum{2}
\includegraphics[scale=0.225]{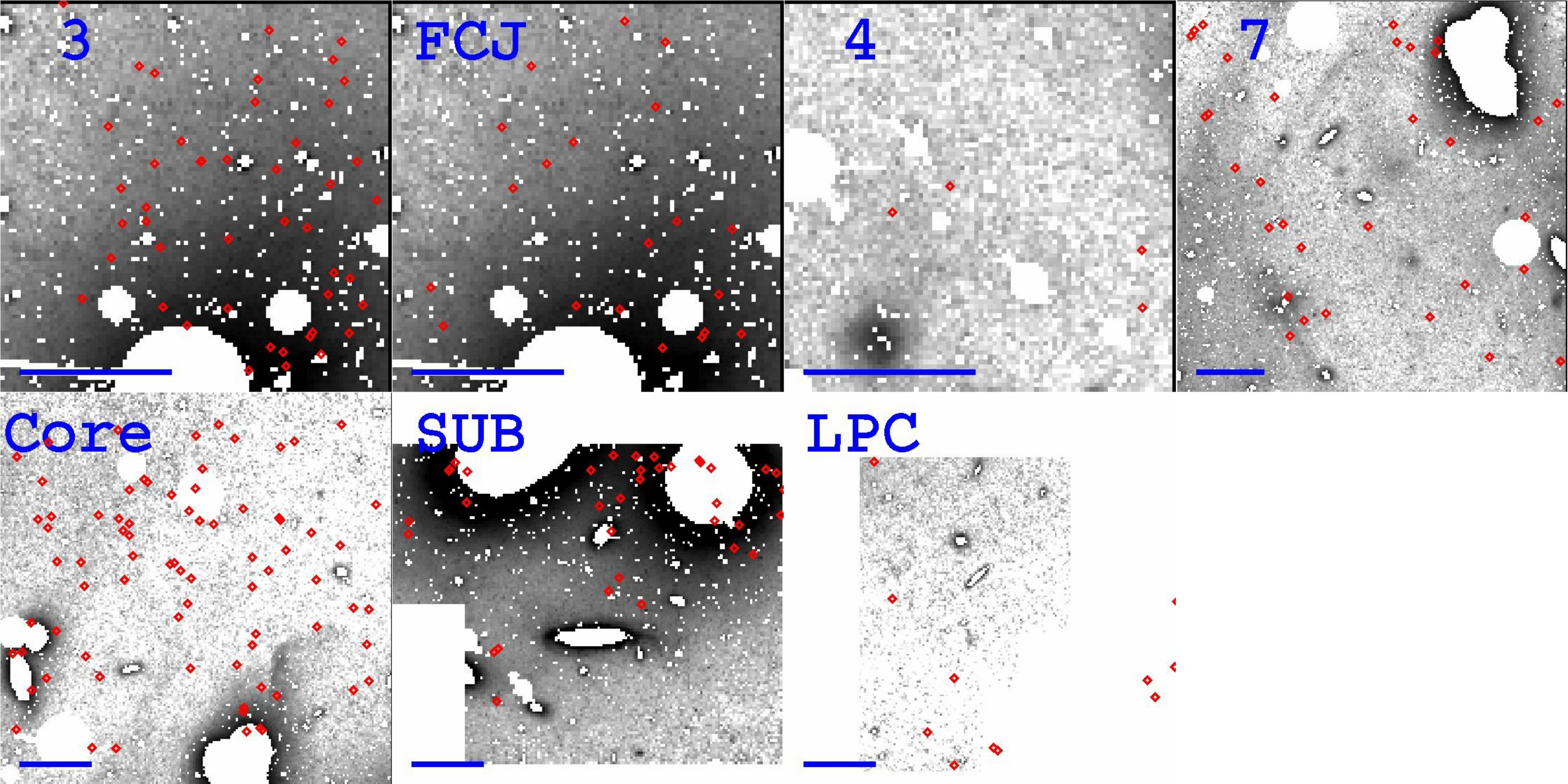}
\caption{IPN candidates from F04 and A05 overlaid on the broadband
  imaging from Mihos \etal (2005). Only candidates brighter than the
  limiting magnitude in each field (see Table 1) are shown. A
  3\arcmin\ scale bar is shown in the lower left of each
  field. The inner regions of galaxies are masked at \muv$<$25
  \magsec\ for all fields except SUB, which is masked at \muv$<$24
  \magsec.
\label{fig:PNeOverlay}}
\end{figure*}

\begin{figure*} \figurenum{3}
\includegraphics[scale=0.225]{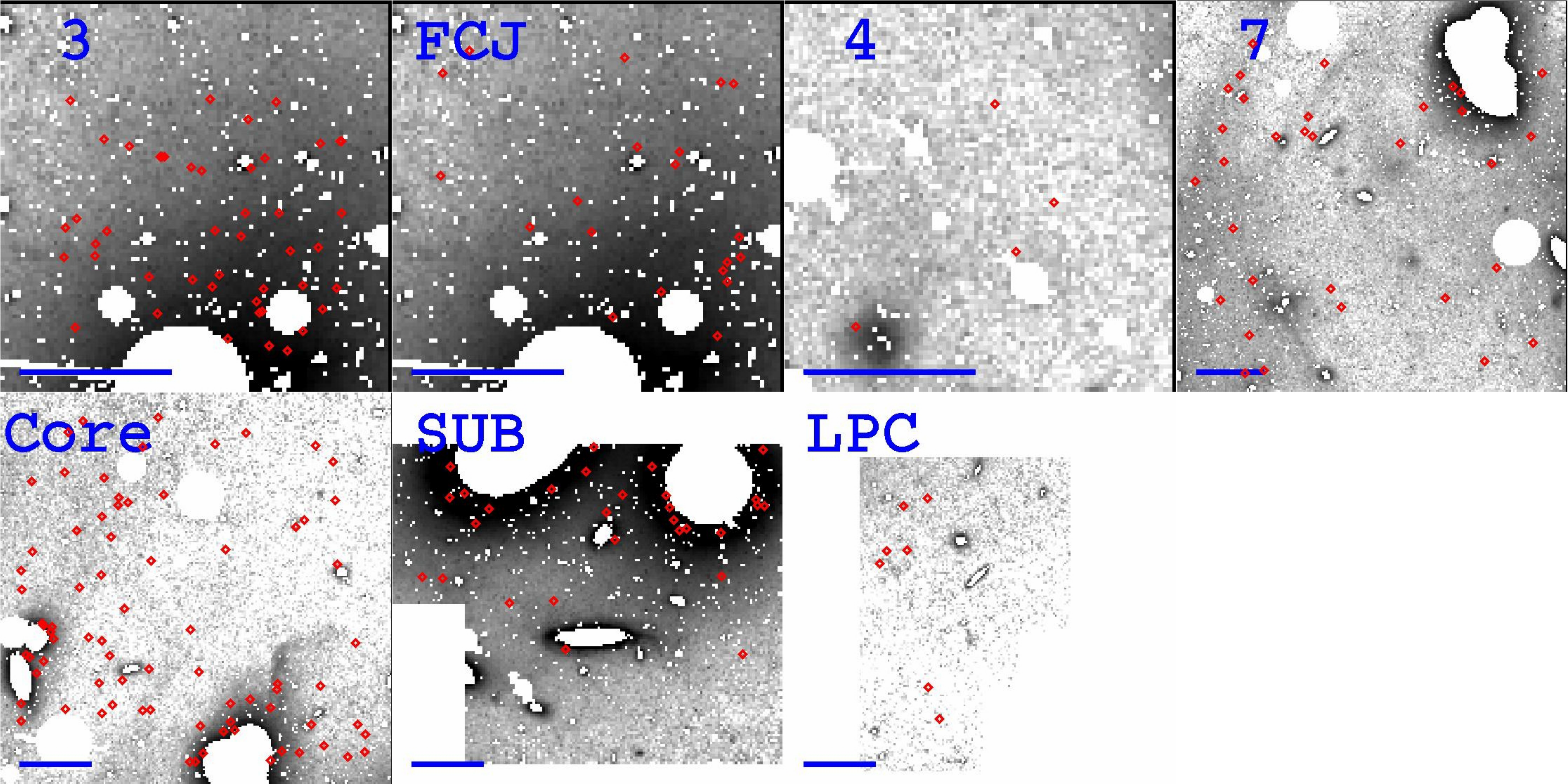}
\caption{Mock IPN catalogs overlaid on the broadband
  imaging from Mihos \etal (2005). The catalogs are generated under
  the assumption that PNe follow the broad band luminosity; see text
  for details. A 3\arcmin\ scale bar is given at the bottom left of
  each field.The inner regions of galaxies are masked at \muv$<$25
  \magsec for all fields except SUB, which is masked at \muv$<$24
  \magsec.
\label{fig:PNeOverlaySim}}
\end{figure*}

\begin{figure} \figurenum{4}
\plotone{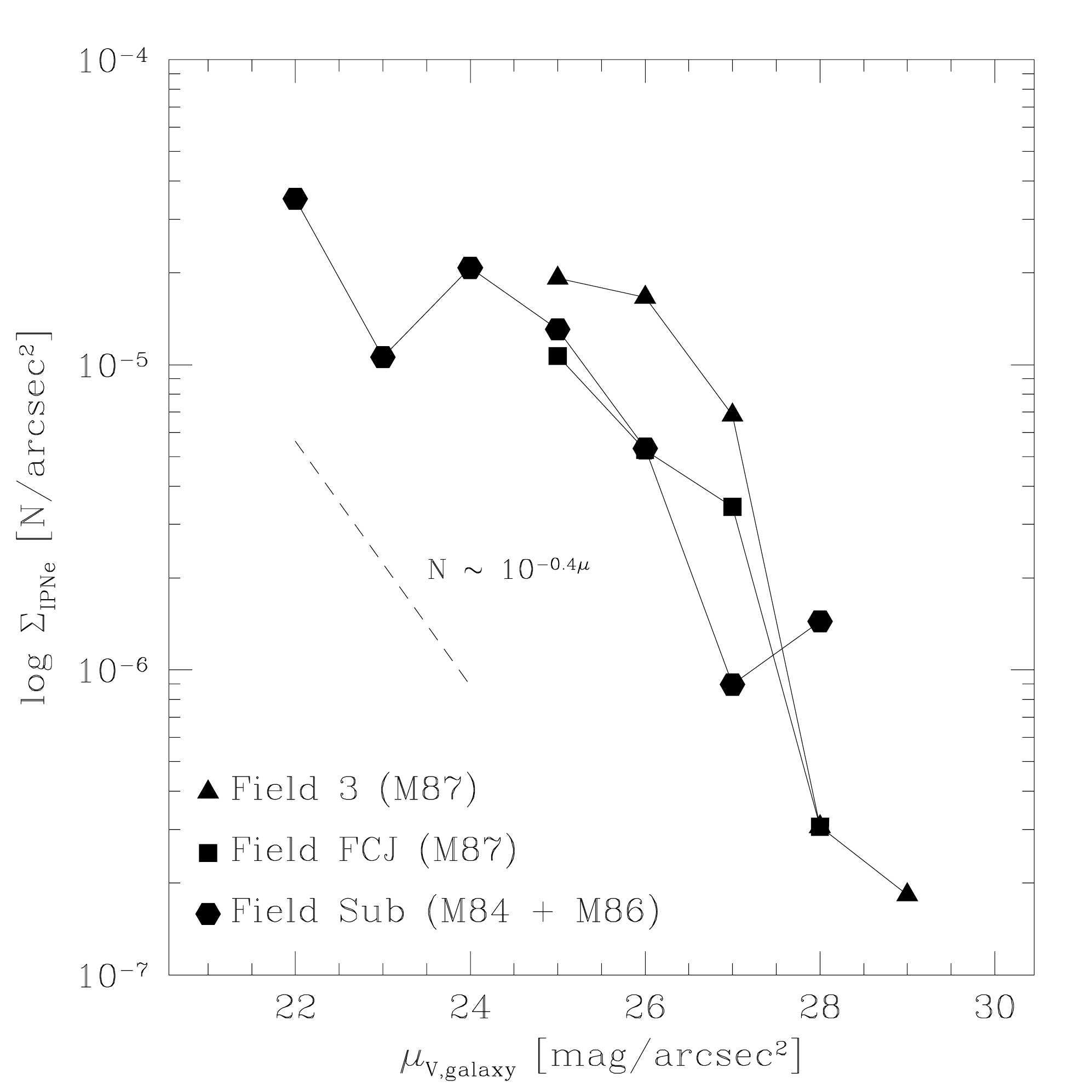}
\caption{
IPN surface density versus modeled galaxy surface brightness near the
ellipticals M87, M86, and M84. The surface brightness models are
isophotal fits to the broad band imaging by Janowiecki \etal (in preparation).
The dashed line shows the expectation if IPN surface density follows
broadband surface brightness. The different limiting magnitudes and
selection criteria for the different IPN datasets may lead to vertical
offsets between the different curves, but the general trend that IPN
surface density traces the light in the halos of the giant ellipticals
seems clear.
\label{fig:findmu}}
\end{figure}

\section{Large Scale Averages: Absolute Measurements of ICL}

We begin with an absolute comparison of the ICL measured in our deep
broadband imaging to that inferred from the IPN measurements by F04
and A05 over larger scales (60--150 kpc).  As discussed in \S2, given
the differences in the various IPN survey fields and catalog
extraction techniques, combining the various IPN datasets is rather
problematic.  Rather than attempting to create a master IPN catalog,
we adopt the analyses of F04 and A05, which convert IPN density to
inferred broad-band surface brightness using statistical corrections
for incompleteness and contamination which are appropriate to the
individual survey fields. We also adopt a value of $\alpha$
(\alphadef) for the conversion which is standardized across all IPN
fields (see below). This then gives us a measure of the absolute level
of ICL in each survey field, which we can then compare to our own
broadband imaging. We also note that the variation between the
analysis by F04 and A05 of Field 3/FCJ can then be viewed as an
estimate of the systematic uncertainty in the IPN-inferred surface
brightness.

As mentioned previously, one needs to adopt a value for the parameter
$\alpha$ to convert IPN counts to broadband surface brightness. The
exact value of $\alpha$ depends on how far down the planetary nebular
luminosity function (PNLF) one detects PNe, as well as the stellar
population mix of the ICL (Buzzoni \etal 2006): more luminous,
metal-rich galaxies have lower values of $\alpha$ than do low
luminosity metal-poor systems (Ciardullo \etal 2005). For simplicity,
we adopt IPN-inferred surface brightnesses calculated using the value
of $\alpha$ determined from studies of individual intracluster stars
in Virgo by Durrell \etal (2004), $\alpha_{2.5} = 23^{+10}_{-12}
\times 10^{-9}$ PNe/L$\sun$, where the ``2.5'' refers to the number of
PNe 2.5 magnitudes below M*, the high-luminosity cutoff in the PNLF
($M*=-4.51$; Ciardullo \etal 2005). This relatively high value of
$\alpha$ measured by Durrell \etal (2004) for the Virgo ICL is
reflective of low luminosity systems ($M_B > -20.5$), but is measured
only over a small field ($\sim$ 4 arcmin$^2$) halfway between M87 and
M86. Over the larger fields surveyed by the IPN surveys and our deep
imaging, there may be significant and systematic differences in the
stellar populations -- and therefore the value of $\alpha$ -- in
Virgo's ICL.  We return to the uncertainty in $\alpha$ in \S 5 below.

We do make one more adjustment to the A05 analysis. A05 quote their
inferred surface brightnesses in \mub, while both our imaging and the
F04 results work in \muv. We therefore transform the A05 results from
\mub\ to \muv\ by adopting a B-V color of 1.2, consistent with the
choice of bolometric correction (BC=$-0.8$) adopted for the stellar
populations in their work.

\begin{figure*} \figurenum{5}
\plotone{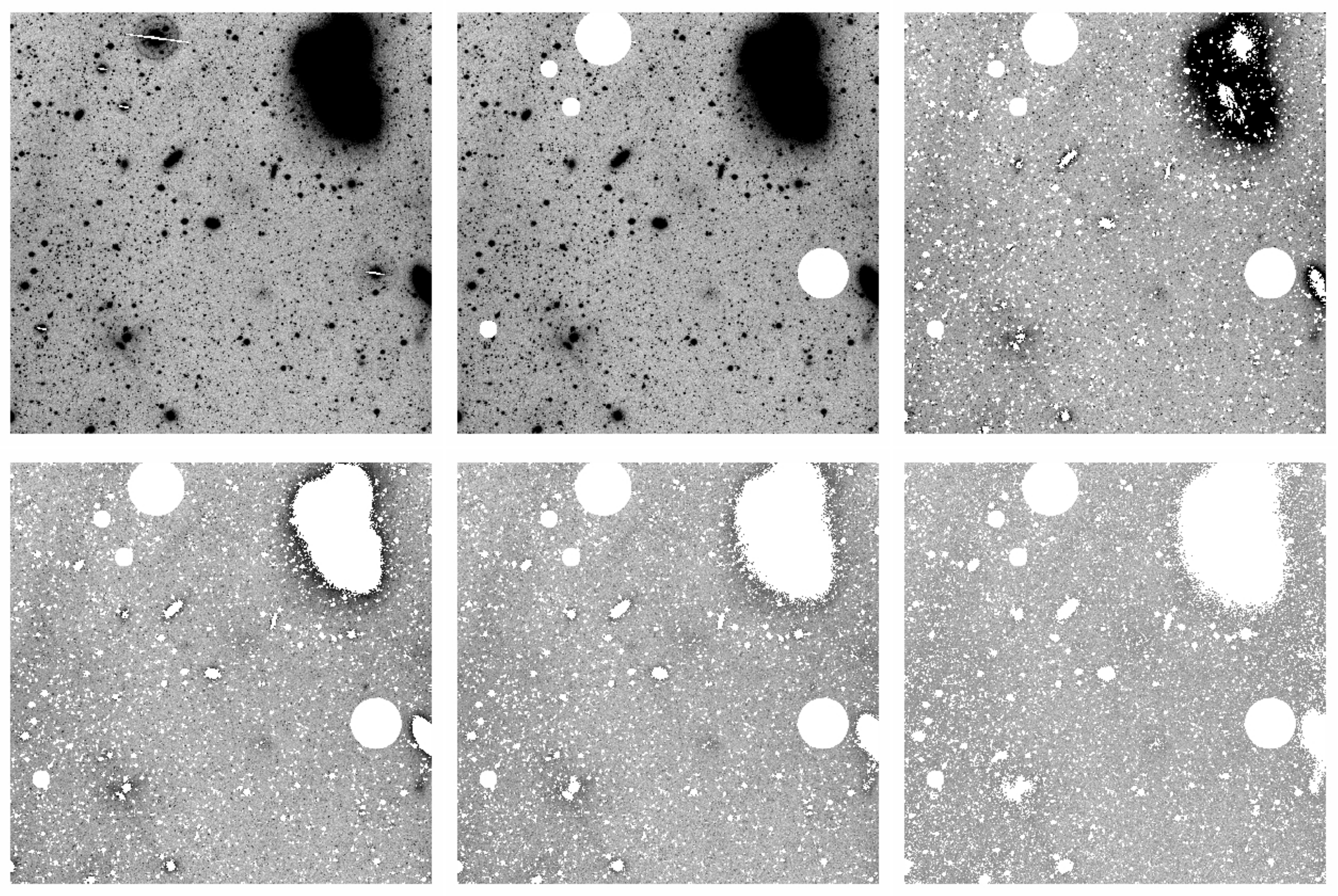}
\caption{The masking process applied to Field 7. The raw image is shown
  in the top left field. The top middle image show the image with the
  bright star mask. The SExtractor mask is added in the top right
  image. The bottom row shows the effects of adding a surface
  brightness mask, masking all pixels brighter than \muv=25.5 (left),
  \muv=26 (middle), and \muv=26.5 (right).
\label{fig2}}
\end{figure*}

To measure the diffuse intracluster light in our fields, we must first
mask out foreground stars and background galaxies (large, bright Virgo
galaxies are handled in a separate step, described below). We mask
using the prescription described in Feldmeier \etal (2002), and refer
the reader there for more details; here we will walk briefly through
the steps, using the Field 7 to illustrate the process (Figure 5). In
brief, we start by manually masking all extremely bright stars, as
well as their diffraction spikes and column bleeds (Figure 5b). We
then run a ring median on the image and subtract off the ring median
image from the raw image; producing an image where the large galaxies
have essentially been removed. We use SExtractor (Bertin \& Arnouts
1996) to identify sources on this median-subtracted image, using a
very aggressive search: all objects 0.6$\sigma$ above sky and at least
3 pixels in size, creating a preliminary mask catalog which consists
of both real objects and faint positive noise spikes.

We next need to filter the noise spikes out of the mask catalog, since
masking only positive noise spikes, but not the corresponding negative
noise spikes, would bias our final measurement of the ICL to
artificially low values. We do this masking in a statistical sense, by
multiplying the median-subtracted image by $-1$ to create a ``negative
image'' and rerunning SExtractor with the same search parameters --
the negative noise features it identifies should have the same
distribution as the positive noise features in the original image, and
we can use this distribution to statistically remove the noise
features from the mask catalog. Specifically, as a function of source
magnitude, we define a ``noise fraction'' (simply the ratio of the
source counts detected in the negative image to those in the actual
image) and then, for each source in the actual image, we use rejection
sampling (based on the magnitude-dependent noise fraction) to
determine whether or not to keep it in our final SExtractor masking
catalog. We add this mask to the bright star mask to produce the
second level of masking (Figure 5c).

The SExtractor mask does a good job of masking small galaxies and
non-saturated stars, but as it is generated from a median-subtracted
image, it fails to capture the large galaxies which disappear during a
median subtraction. It also spuriously identifies as discrete sources
bright spots (either real or noise) in the inner regions of galaxies.
To overcome this, we add one last layer of masking, which masks out
the high surface brightness regions of galaxies. This mask also
qualitatively captures an areal selection effect wherein IPN studies
generally do not concentrate on the inner regions of galaxies where
the IPN detection is difficult against the high surface brightness of
the galaxy.

To create this final mask, we mask all pixels brighter than a given
surface brightness threshold, taken to vary from \muv=25.5 to
\muv=26.5. Figure 5d-5f show the mask at these varying surface
brightness levels; at \muv=25.5, light is clearly ``leaking out'' from
the large galaxies, while at \muv=26.5, the mask appears too
aggressive, blocking out both diffuse light (for example, the faint
tidal features extending south from NGC 4425 near bottom right) and
individual noise spikes. While our preference from an examination of
these masks is to place most emphasis on the results using the \muv=26
mask, we use the results of the analysis using the \muv=25.5 and
\muv=26.5 masks as a conservative estimate of our uncertainty.

Given the masked image, we then define the photometric surface
brightness of the diffuse light in each field to be the mean surface
brightness of the unmasked pixels below the surface brightness
threshold. Since IPNe are expected to be an unbiased tracer of
luminosity, this simple definition of the ICL surface brightness
should connect to the IPN-inferred surface brightness measurements
most directly. In Figure 6 we plot a comparison of the photometric and
IPN-inferred ICL surface brightness measurements for the six fields
shown in Figure 1.

An examination of Figure 6 shows a rough correspondence between the
two measures of ICL. Field 3/FCJ shows the highest inferred surface
brightnesses using both techniques, and it is clear from Figure 1 that
this field largely covers the extended stellar halo of M87. A large
fraction of the PNe in this field are likely associated with M87
rather than being true intracluster objects. Indeed, followup
spectroscopy of 15 IPN candidates in the FCJ field by Arnaboldi \etal
(2004) shows that 12 of the 15 candidates have a narrow velocity
distribution ($\sigma=247 \pm 52$ km/s) that can be associated with
M87's halo.

Field SUB also has significant overlap with the halos of bright
galaxies, most notably the ellipticals M84 and M86. In this field,
many of the PNe candidates lie close to bright galaxies and have a
velocity structure which is closely correlated with the galaxy
velocities (Arnaboldi \etal 2004), again suggesting that we are not
probing a true intracluster population here.

Fields 7, 4, and Core all have lower photometrically-inferred surface
brightnesses, and correspond to regions which have more intracluster
area. The bright interacting pair NGC 4435/38 falls in both Field 7
and Field Core, but covers a relatively small fraction of the total
area of these fields. The smaller Field 4 contains one low surface
brightness dwarf (VCC 1052; Bingelli, Sandage, \& Tammann 1985), and
no other galaxies of note.

In both surface brightness and IPNe, the LPC field has the faintest
inferred intracluster light. Lying south of the M84/M86 complex, this
field contains no bright galaxies and overlaps only minimally with the
extended diffuse light around M84/M86. This field also shows the
largest disagreement between the photometric and IPN-inferred surface
brightness, -- our measurement is approximately 3.5 mag/arcsec$^2$
brighter than that inferred from the IPN data. As noted previously,
the IPN candidates in this field are also found largely to the
southwest portion of the field, away from the complex of galaxies just
north of the field. So while LPC has the least amount of ICL measured
both by broad band light and IPN studies, the marked disagreement in
both the absolute level of diffuse light and its spatial distribution
makes the comparison between the two methods less than satisfactory.
One caveat is that our broad-band imaging only partially covers the
full LPC field, such that we are biased towards the high surface
brightness portion of the field closer to the luminous
ellipticals. However, if we presume that there is {\it no} ICL present
in the portion of the field we do not image, the net surface
brightness we would measure for the entire field would drop only by
0.5 mag/arcsec$^2$, still leaving us with a significant discrepancy.

An examination of Figure 6 shows a tendency for the IPN-inferred
surface brightness to be systematically fainter than the photometric
surface brightness. Again, the IPN-inferred surface brightness is
based on the Durrell \etal (2004) value of $\alpha_{2.5}=23 \times
10^{-9}$, and a systematic offset between the two surface brightnesses
could come from the use of an incorrect value for $\alpha$. If instead
we demand that the data scatter around a line of equality and {\it
fit} for a value of $\alpha_{2.5}$, we get $\alpha_{2.5}=(13.8 \pm
4.2) \times 10^{-9}$, lower than the Durrell \etal value. Such values
of $\alpha_{2.5}$ are similar to those measured in luminous
ellipticals (Ciardullo \etal 2005), and may suggest an ICL which
originates from stellar populations stripped from Virgo's luminous
ellipticals. However, the scatter in the relationship shown is quite
large, and the systematic effects present in both datasets (see \S4)
complicate this simple analysis.

\begin{figure} \figurenum{6}
\plotone{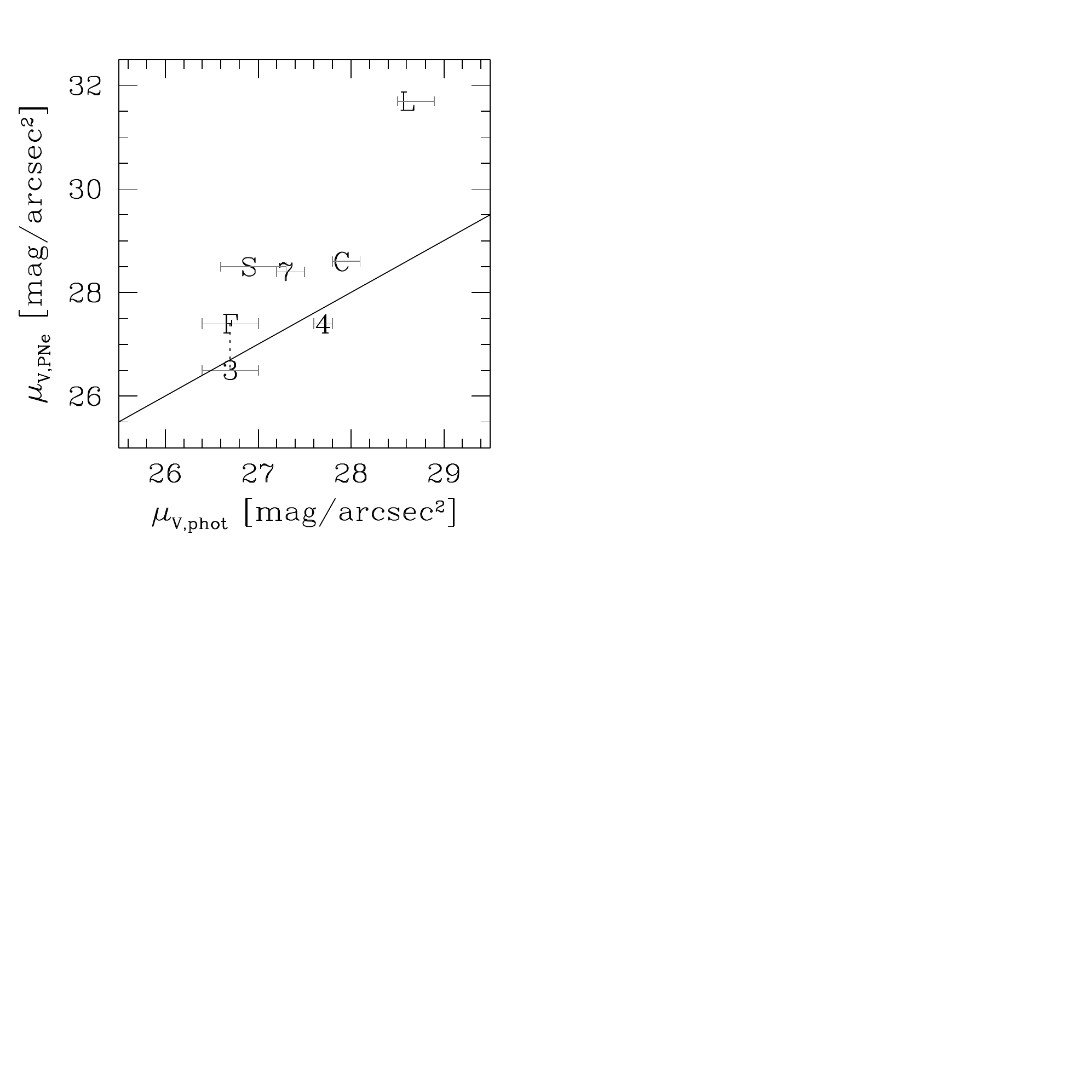}
\caption{Comparison of photometric surface brightness with that
  inferred from IPN counts. The vertical axis shows the IPN-inferred
  surface brightnesses (from F04 and A05), while the horizontal axis
  plots the surface brightness measured in our broadband imaging
  (Mihos \etal 2005) using a data mask of $\mu_{V,mask}=26$. The
  plotted errorbars are not photometric uncertainties, but rather
  represent the range of measured surface brightness for masks of
  $\mu_{V,mask}=25.5$ and $\mu_{V,mask}=26.5$. The symbols refer to the
  field names, and a line of equality is shown; this is not a
  fit. Below the equality line, the IPN-inferred surface brightnesses
  are brighter; above the line the broad band measurements are
  brighter. A dotted line connects Fields FCJ and 3, as an indicator
  that these represent independent IPN analyses of the same field.
\label{fig3}}
\end{figure}

\section{Small-scale Correlations: Relative Measurements of Excess
  Luminosity around Individual IPN Candidates}

The previous section looked at the correlation between the absolute
levels of ICL measured by broad band imaging and IPN surveys on the
$\sim$ 100 kpc scales. However, measuring the correlation between ICL
and IPNe on these scales blurs together different environments -- an
individual IPN field may simultaneously contain the outskirts of
luminous galaxies, faint tidal streams, and true intracluster
space. The variation in environment may make it hard to identify a
clean connection between the diffuse light and the IPNe, and so we
wish also to probe the connection on smaller scales ($<$ 15 kpc),
where individual galaxies or tidal streams can be isolated.

To examine the correlation between IPN candidates and diffuse
broadband light on these smaller scales, we look {\it within}
individual survey fields, and measure the amount of {\it excess}
luminosity around IPN candidates (compared to random positions in the
field). If the IPN candidates trace the broadband luminosity, this
should manifest as a luminosity excess around the candidates, and
since it is a relative measurement, it is more robust against
systematics of sky subtraction at low surface brightness, where we are
most interested in testing the correlation.

Our quantitative technique explores the luminosity excess in boxes of
varying angular scale that contain one or more IPN candidates,
compared to randomly scattered boxes.  For a field with \ncand IPN
candidates, we randomly sample the field using boxes of a given size,
until we have \ncand boxes each containing at least one IPN candidate;
the median pixel intensity $I_1$ (in ADU) is then calculated using
{\it all} unmasked pixels in these \ncand boxes. An identical
procedure is used to measure the median pixel intensity in \ncand
randomly placed boxes ($I_r$). We then define the value $\Delta_{1r}$
to be the difference in median pixel intensity between these two
samples: $\Delta_{1r}=I_1 - I_r$.  We repeat this process 40 times,
and calculate the median and upper and lower quartile of the
$\Delta$'s over these 40 trials.  For these trials, we choose sampling
boxes which range in size from 3x3 pixels (4.35\arcsec or 0.3 kpc on a
side) to 131x131 pixels (190\arcsec or 14.6 kpc on a side) to look for
correlations over a range of physical scales.\footnote{We note that in
a relative measurement like this, even if a correlation between IPNe
and diffuse light exists, the relative excess of light around IPN
candidates (compared to the field in general) will trend toward zero
as larger and larger box sizes are sampled, even if there is
significant ICL which contributes to a high absolute measurement of
surface brightness for the field as a whole (compared to other
fields).}  In this analysis, we also restrict ourselves only to IPN
candidates brighter than the limiting magnitude of each field given by
F04 and A05.

One concern with such an analysis is a selection effect for detecting
IPN candidates: individual PNe will be hard to distinguish against a
high surface brightness background such as the inner regions of bright
galaxies. This places a limiting surface brightness above which IPN
candidates will not be detectable. The exact surface brightness limit
will depend on the observational details and analysis of the different
IPN surveys and are hard to uniquely determine. Nonetheless, we can
mimic such a selection function by masking all pixels brighter than a
certain limiting magnitude in our imaging data before doing our
analysis. An examination of Figure 2 shows that for most fields except
SUB, nearly all IPN candidates are found at surface brightnesses \muv
$>$ 25 \magsec; for SUB, this surface brightness limit is a magnitude
brighter, \muv $>$ 24 \magsec. Accordingly, we apply surface
brightness masks to the data at these levels to mimic the IPN surface
brightness selection effect.

\begin{figure*} \figurenum{7}
\includegraphics[scale=0.85]{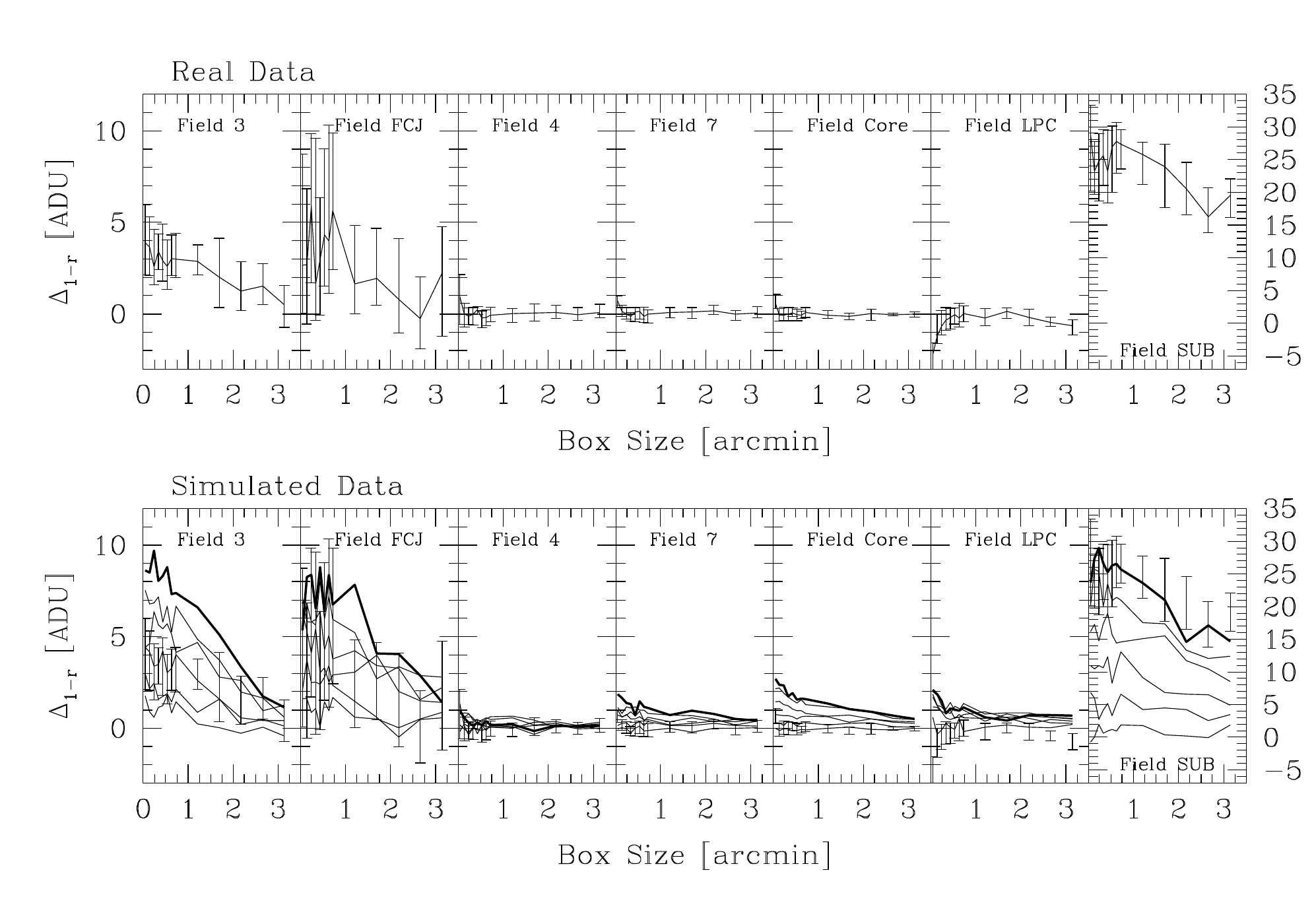}
\caption{
Top: Luminosity excess (in ADU) versus box size for randomly sampled boxes
chosen to contain IPN candidates, compared to purely randomly placed
boxes. Bottom: Same as above, but for {\it mock} IPN catalogs
constructed using a ``PNe-trace-light'' model. In the bottom panels,
comparisons with the pure mock catalogs are shown using a bold line,
while the lighter lines are for catalogs contaminated by randomly
distributed contaminants at the 20\%, 40\%, 60\%, 80\%, and 100\%
level. For comparison, the actual correlation signal from the real
data shown in top panel is reproduced in the bottom panel. See text
for  further details. In both panels, the vertical scale for Field SUB
is different from the others; the proper scale for SUB is shown at the
right of the figure.
\label{fig:fielddeltas}}
\end{figure*}

The correlations for each field are shown in Figure 7a, where the
errorbars represent the upper and lower quartile of the measurements.
We see the strongest correlations in the fields near M87 (Fields 3 and
FCJ) and M86/M84 (Field SUB), which have on average higher surface
brightness. Near M87, the luminosity excess is stronger on small
scales, and slowly declines towards zero on larger scales. This trend
is seen in both Field 3 and FCJ, although the significance is lower in
field FCJ due to the smaller number of IPN candidates detected in the
analysis by A05. Clearly in this field, the IPN candidates are tracing
M87's diffuse halo, and indeed may not be truly {\it intracluster}
objects, but reflect instead the bound stellar population around
M87. As mentioned in \S3, this suggestion is borne out by the
kinematic study of IPNe in this field by Arnaboldi \etal (2004), which
reveal a velocity distribution to the IPNe which is similar to M87
itself.  In Field SUB, which contains M86, M84, and several other
luminous galaxies, the correlation is even stronger than in Field
3/FCJ, both in amplitude (since IPNe are found deeper inside galaxies
in this field) and in the decline with angular scale. In this field,
the IPN candidates are almost all found near M86 and M84, and again
are likely to be associated with the giant ellipticals themselves,
rather than being intracluster objects.

In the fields further away from large galaxies (Fields 4, 7, Core, and
LPC), where the average surface brightness is much lower, we find no
correlation between the local broadband surface brightness and the IPN
candidates at any spatial scale. On the smallest scale, it is possible
in principle to detect the IPN candidates themselves (the brightest
PNe would have $m_{5007}=26.0$, yielding a peak intensity of $\sim$
0.9 ADU in our image), but we see no evidence of this in the
data. There are small fluctuations ($< 1$ ADU) at these scales in some
fields, but we find no evidence of point sources when we co-add
the images around the location of the IPN candidates. This fact,
combined with the fact that the signal does not depend on the number
of IPN candidates, and is in fact negative in Field LPC, leads us to
believe that fluctuations on these scales are simply due to
statistical noise.

In the case of Field LPC, we actually see a flux {\it deficit} around
IPN candidates. While there are not many IPN candidates in this field,
this flux deficit does appear to be statistically meaningful. This
again is a manifestation of light and IPN candidates being
anti-correlated in this field: randomly placed boxes sample the broad
band light in an unbiased fashion, while boxes that contain IPN
candidates are biased to the southern portion of the field, away from
the diffuse light surrounding the complex of galaxies near (and
including) M86 and M84. These differences lead to the anti-correlation
seen in the LPC field.

In summary, we see correlations between the IPN candidates and the
broadband light in higher surface brightness regions near the luminous
ellipticals in Virgo, but in true intracluster fields at low surface
brightness there seems to be no small-scale correlation visible at
all, save for the anti-correlation seen in LPC.  One concern with this
analysis remains the very different areas and surface densities of IPN
candidates in the different fields.  Given the relatively small number
of IPN candidates, and the likely contamination of the candidate list
by background emission line galaxies, is it likely that we would able
to {\it detect} the correlation between broadband light and IPN
candidates in these datasets? We address this question by creating an
artificial sample of IPN candidates from the broad band image, under
the explicit assumption that PNe trace light (as described in \S2),
and that contaminating objects are uniformly distributed across the
field.  Then, for a given contamination fraction $f$, we randomly draw
$(1-f)N_c$ PNe from the catalog (where $N_c$ is the number of
candidate IPNe actually observed in each field), and add another
$fN_c$ PNe randomly distributed across the image. We then analyze this
artificial dataset in the same way that the real candidate lists were
analyzed, to see what correlation signal we would detect. For each of
the 40 trials we select a new random sample of IPNe from the master
list, so that our results are not simply reflecting the distribution
of a single realization of modeled IPNe.

An example of a single realization of IPN candidates under the
PNe-follows-light model (with no contaminant) was shown in Figure 3,
while the statistical analysis of the modeled IPN datasets is shown in
Figure 7b. For each field we show curves for the artificial datasets
given contamination fractions of 0\%, 20\%, 40\%, 60\%, 80\%, and
100\%, with the true (observed) correlations overlaid. As expected,
contamination acts to reduce the correlation signal quite quickly. The
only field consistent with no contamination is Field SUB, which is not
surprising since its candidate IPNe were detected using the two-line
method which significantly reduces contamination. In Field 3 and FCJ,
the data is consistent with a model where IPNe follow light (in this
case, the extended halo of M87), assuming a contamination fraction of
$\sim$ 40\%, somewhat higher than the value of 20\% adopted by F04 and
reported by Arnaboldi \etal (2004) from follow-up spectroscopy of a
subset of candidates.

In Fields 7 and LPC, the model shows that a small signal might be
detectable, but at a very low significance level that is quickly
washed out by even a modest amount of contamination. Given the
contamination that is known to exist in these fields, it is no
surprise that we see no correlation, although the {\it
anti-correlation} seen in Field LPC is still unexplained. With more
IPN candidates, the correlation signal in Core should have been
detectable if the contamination was moderate; unfortunately,
spectroscopic follow-up has shown a very low confirmation rate for the
IPN candidates (32\%; Arnaboldi \etal 2004). If the contamination rate
is as high as 70\%, the correlation signal in this field is clearly
undetectable. And finally, as expected for Field 4, there are simply
not enough IPN candidates to distinguish any correlation -- we show
the results for this field simply for completeness' sake.

Analyzing the mock IPN catalog moderates somewhat the disconnect seen
between the broad band light and the IPN candidate catalogs. Where we
expect a strong signal and modest contamination -- i.e., near the
giant ellipticals -- we do in fact recover the correlation (Fields 3,
FCJ, and SUB). In the intracluster fields, the signal is much weaker
and contamination acts to wash out the correlation.

\section{Discussion}

Since PNe are formed in the late stages of the evolution of low mass
stars, they should trace the stellar mass of galaxies. In galaxy
clusters, this means that IPNe should, in principle, trace the diffuse
intracluster light formed from stars stripped from their host
galaxies. We have compared the ICL detected in broad band light with
candidate IPNe detected in narrowband imaging surveys, and find only a
rather rough correspondence between the two. The strength or absence
of the correlation clearly depends on local surface brightness. At
higher surface brightness, near the luminous ellipticals M87, M86, and
M84, the correspondence is strong, as evidenced by Fields 3, FCJ, and
Sub. In truly intracluster fields where the average surface brightness
is lower, however, the correlation is much weaker, on both large
($\sim$ 100 kpc) and small ($\sim$ 1-10 kpc) scales. Unfortunately,
while the expectation that IPNe trace light is well-motivated, many
effects, both systematic and statistical, can act to cloud the
connection. We consider each in turn now.

\bigskip
\noindent{\sl IPN Catalog Contamination}

The biggest complication to the analysis is that of contamination in
the IPN catalogs, as described extensively in the previous
sections. In the field-averaged absolute comparison detailed in \S3, the
IPN-inferred surface brightnesses taken from F04 and A05 factor in
this contamination statistically. Clearly, however, this correction
can be significant and can vary both from field to field (due to
differences in limiting magnitude and off-band subtraction; see F04
and A05 for details) and likely within individual fields due to
clustering of background sources (\eg Ciardullo \etal 2002).

Contamination is even more problematic when correlating the light on
small scales with individual IPN candidates, as was done in \S4. Our
Monte-Carlo modeling of the effects of contamination shows that
contamination fractions of $\sim$ 40\% are needed to explain
discrepancy between the data and the ``PNe-follow-light'' model, a
number consistent with the rates of contamination inferred from the
extant spectroscopic followup that exists for several of the fields
(see the discussion in \S2.2 and in Arnaboldi \etal 2004). The only
field consistent with the pure ``PNe-follow-light' model is Field SUB,
and the IPN catalog for this field was generated using the two-line
method (Okamura \etal 2002), which significantly cuts out
contamination from background sources and photometric errors.

We also note also that a population of contaminants (real or artificial)
that is distributely uniformly across the fields could easily give
rise to the fact that the correlation we detect is a strong function
of surface brightness. If PNe do follow light, a constant density of
contaminants means that the contamination fraction will increase at lower
surface brightness, leading to a correlation at high surface
brightness that washes out as lower surface brightness regions are probed.

\bigskip
\noindent{\sl The Effects of Stellar Populations}

Another possible complication is variation in the parameter
$\alpha(=N_{PNe}/L_{bol})$, which converts PNe number density to
broadband surface brightness. This parameter is sensitive to the
underlying stellar population (Buzzoni \etal 2006); measurements of
$\alpha$ within individual galaxies show a scatter of a factor of five
between systems, with some suggestion that the scatter is linked to
the metallicity of the host system (Ciardullo \etal 2005). In
converting the PNe measurements to equivalent V-band surface
brightness, we have adopted the value of $\alpha$ obtained by Durrell
\etal (2004) which is typical of low luminosity systems. On average,
the IPN-inferred surface brightnesses are fainter than the
photometrically derived surface brightnesses by about one
mag/arcsec$^2$, a discrepancy would could be accounted for by adopting
a value of $\alpha_{2.5}=9$ instead of the Durrell \etal (2004) value
of $\alpha_{2.5}=23$. This would suggest that the ICL in the Virgo
fields may arise from material associated with or stripped from
Virgo's giant ellipticals. Indeed, three of the fields lie close to
the luminous ellipticals M87, M86, and M84 (\ie the Subaru and 3/FCJ
fields), consistent with this idea.

More globally in Virgo, however, the exact value of $\alpha$ could
show systematic and complex variations depending on the progenitor
population of galaxies contributing to the ICL, and it may be naive to
expect one value of $\alpha$ to accurately describe the entirety of
the Virgo ICL population. One might hope with sufficient broadband
data and uncontaminated IPN catalogs to invert the problem and map out
the variations in ICL stellar populations across the cluster, but
clearly the existing data shown here is not sufficient for such a
task.

The stochastic nature of the IPN population statistics also affects
the detected correlations.  The number of IPNe per unit stellar
luminosity (given by $\alpha$) is low, such that quantitative
measurements will be affected by the Poisson noise in the number of
detected IPNe. For example, adopting the value of $\alpha$ measured
for the ICL in Virgo by Durrell \etal (2004) and a color of $B-V=1$
for the ICL, the tidal stream emanating from M87 should only have
$\sim$ 5 PNe within 1 magnitude of M* spread over its 100 kpc
extent. Within the fields analyzed here, there are between 10 and 100
candidate IPNe per field, but given the high contamination fraction in
some fields, the expected number of {\it true} IPNe is more like a few
to 50 per field. With such small numbers of IPNe, the statistical
fluctuations due to Poisson noise in the number counts will be large.

\bigskip
\noindent{\sl IPN Spatial Distribution}

One potentially important systematic difference between broadband
imaging surveys for diffuse ICL and searches for IPNe relates to the
line-of-sight depth of the ICL. As surface brightness is a
distance-independent quantity, our deep imaging will detect ICL over
the entire depth of the Virgo cluster. As the IPN surveys detect point
source objects down to a given limiting magnitude, at a certain depth
these surveys will begin to miss IPNe. The Virgo cluster has complex
substructure and studies of its three dimensional structure have
suggested significant depth. While much of this structure is on scales
larger than the so-called Virgo ``Cloud A'' which our data probes,
even the core of Virgo has evidence of substructure in its depth. For
example, Jerjen \etal 2003 use SBF distances to early-type galaxies to
show that the distance distribution is broad, with a front-to-back
depth for the cluster of 5.6 Mpc (defined as $\pm 2\sigma$ in the
line-of-sight distances). Moreover, the distribution is bimodal, with
peaks at 16 and 18.5 Mpc associated with the M87 and M86 subclusters,
respectively. More recently, Mei \etal (2007) measure a smaller depth
of the Virgo core of 2.4 $\pm$ 0.4 Mpc. However, the depth of the
cluster defined by the spiral population may be much larger, as the
spiral population in Virgo is more spatially extended than is the
early-type galaxy population (Binggeli, Tammann, \& Sandage
1987). Indeed, while Cepheid distances to Virgo spirals cluster around
a central value of 15 Mpc (Freedman \etal 2001), outliers do exist, as
shown by Cepheid distance to NGC 4639 (d=25 Mpc; Saha \etal 1997).

The limiting depth of the IPN surveys depends both on the survey
limiting magnitude and the luminosity of the IPNe. The planetary
nebula luminosity function shows a sharp break at $M^*_{5007}=-4.51$
(Ciardullo \etal 2002); very few PNe are found brighter than this
value. At a canonical Virgo distance of 16 Mpc, this corresponds to an
apparent magnitude of $m^*_{5007}=26.5$, such that all surveys can
detect the brightest IPN candidates at least to the distance of
Virgo. However, because of the sharp cutoff in the PNLF at $M^*$, one
needs to go at least half a magnitude down the luminosity function to
get appreciable samples of IPNe. As a result, the limiting magnitude
of several of the IPN fields means that only IPNe in the {\it near
side} of the cluster will be detected. Table 1 shows the limiting
$M^*$ depth of each field; going half a magnitude fainter will reduce
this depth by 20\%. The problem is particularly severe for Fields 4
and 7, which have the shallowest limiting depth. Other fields have
gone deeper and should detect IPNe further along the line of sight
through the cluster, but all will be systematically biased towards
detecting IPNe on the near side of Virgo.

In contrast to the IPN surveys, our deep broadband imaging will probe
the full depth of the cluster and therefore can reveal ICL that is too
distant to have detectable IPNe. This effect would result in
photometric surface brightnesses which are systematically {\it
brighter} than those inferred from IPN counts. Again, Figure 6 shows a
slight systematic trend in exactly this sense, but the scatter is
large, and there is no obvious correlation with limiting depth of the
field. For example, the fairly shallow Field 4 shows agreement between
the two methods, while the deeper SUB and LPC fields show large
discrepancies. Deeper IPN imaging will be needed to test this issue
further.

\bigskip
\noindent{\sl Other Methodological Uncertainties}

Certainly the uncertainties on these measurements are large; both
techniques push the limits of the available data, and there are large
and differing systematic uncertainties in each technique. Indeed, even
{\it within} each technique the uncertainties are large. For example,
depending on the adopted surface brightness threshold for masking, our
data yield ICL surface brightnesses which can vary by $\sim$ 1
mag/arcsec$^2$ (see Table 1). Similarly, different analyses of the
{\it same} imaging data for Field 3/FCJ by F04 and A05 -- using
different detection criteria -- yield inferred surface brightnesses
that differ again by $\sim$ 1 mag/arcsec$^2$.

One last systematic effect to consider is that of over-subtraction of
the ICL during the sky subtraction phase of the broadband imaging data
reduction. During this process, we fit planes to the sky and then
subtract the fitted planes; if there is diffuse, structureless ICL
component in the Virgo Cluster core which subtends a size comparable
to the scale of our image ($\sim 1.5\degr$ or 420 kpc on a side), we
will have no way of distinguishing this from the night sky, and it
will be subtracted off. In this way, our broadband ICL estimates could
be systematically low. However, for a cluster with such significant
substructure in its galaxy population, it seems unlikely that a
substantial amount of the ICL would be smoothly distributed over such
a broad scale. Furthermore, the amount of light would be substantial:
spread out over a 420x420 kpc field, a smooth ICL component at $\mu_V
= 27$ would have a total luminosity of $\sim 10^{12} L\sun$, five
times brighter than the total luminosity of M87, M86, and M84 {\it
combined}.  This would skew the ICL fraction of the cluster to
extraordinarily high values, much higher than expected from
simulations of cluster collapse. Summing up the luminosity of all VCC
galaxies (Binggeli \etal 1985) in our imaging field, we obtain a total
blue luminosity of $2.5\times 10^{11} L\sun$; adopting an ICL color of
B-V=1, this would imply a {\it missing} ICL fraction of $\sim$ 60\%,
not factoring in the ICL that we actually do detect in our
imaging. This suggests that if any ICL was removed artificially during
the sky subtraction stage, it must be at much lower surface
brightnesses than we are detecting in our deep imaging.

\bigskip

Given these uncertainties, it is perhaps encouraging that we recover
even the weak correlations discussed in \S3 and \S4. The hope of using
deep surface photometry to act as a finding chart for IPN surveys
remains valid; however, the IPN surveys need to be both deeper (to
uncover more IPNe and improve the Poisson statistics) and rid of
contaminants. With single-line narrow band imaging, these two needs
work at cross purposes, as contamination gets worse at fainter levels
(\eg Ciardullo \etal 2002). Unfortunately, the hope of constraining
ICL stellar populations through a mapping of $\alpha$ across the
cluster is clearly unfeasible without resolving these issues. This
task will not prove easy.

However, there are a number of observationally feasible ways to reduce
the amount of contamination from current and future IPN surveys.
Imaging observations of IPN candidates in better seeing than the
original surveys (the average seeing of all of the IPN fields is
1.25\arcsec) will remove a significant fraction of contaminating
objects.  At distances of $\approx$ 16 Mpc, all IPNe should remain
spatially unresolved, while background contaminating objects such as
bright Ly$\alpha$ galaxies at z = 3.1 have half-light diameters as
large as 0.6\arcsec (Gronwall et al. 2008, in prep).  Therefore,
narrow-band [O~III] imaging with significantly better ambient seeing,
imaging systems with adaptive optics, or imaging systems that use
orthogonal transfer CCDs (Jacoby et al. 2002; Howell et al. 2003) will
be extremely helpful in removing contaminants.

Although Ly$\alpha$ galaxies generally have high observed equivalent
widths ($>$ 80\AA), the density of Ly$\alpha$ galaxies decreases
exponentially as the equivalent width increases (Gronwall et
al. 2007).  Thus, the deeper the off-band imaging is in an IPN survey,
the higher equivalent width limit will be obtained, and the fewer
contaminants will enter the sample.  Similarly, using a third image as
a ``veto'' filter would also reduce the possibility of contamination

All genuine IPNe should also emit in other emission lines, the
brightest of which are the $\lambda$ 4959 [O~III] emission line and
the H$\alpha$ + [N~II] complex.  Unfortunately, both of these emission
line regions are $\approx$ 3 times fainter than the $\lambda$ 5007
[O~III] line (Storey \& Zeippen 2000; Ciardullo et al. 2002), making
confirmatory observations more difficult.  However, it is worth noting
that the SUB field observations used deep H$\alpha$ + [N~II] imaging,
and we have found that this field is consistent with no contamination
(\S 4).

However, most definitive way to completely eliminate contamination of
IPN catalogs is still follow-up spectroscopic observations.  IPNe will
have narrow emission lines compared to contaminating sources (Freeman
et al. 2000; Arnaboldi et al. 2004), the $\lambda$ 4959 [O~III] should
always be present, and there should be no sign of continuum emission.
Despite some significant efforts (Freeman et al. 2000; Arnaboldi et
al. 2004; Arnaboldi et al. 2008; Doherty et al. 2008), the fraction of
IPN candidates that have been spectroscopically observed is still
quite small. With the growing availability of high throughput,
wide-field, multi-object spectrographs on large telescopes, this
situation should improve.  Since the radial velocities of IPNe are of
considerable interest for dynamical studies of the intracluster light,
such follow-up observations will have multiple benefits.

On the broadband imaging side, going deeper not only will prove
challenging technically, but will also quickly run into a fundamental
background: the backscattering of galactic light by dust in the Milky
Way at high galactic latitude. This scattered light shows up in deep
imaging as diffuse light similar in morphology to the diffuse ICL (\eg
Sandage 1976; de Vries \& le Poole 1985; Witt \etal 2008; Janowiecki
\etal in preparation). This high galactic latitude dust can be traced
by its thermal emission in the far infrared, and dust maps constructed
from IRAS imaging (Schlegel, Finkbeiner, \& Davis 1998) show that the
Virgo core fortuitously sits in a hole in the galactic cirrus, such
that the deep surface photometry shown in Figure 1 is relatively
unimpacted by this dust. However, surrounding regions have
significantly more cirrus and detecting and quantifying the ICL in
those regions will be much more difficult. For example, in our imaging
of fields to the north and southeast of the the field shown in Figure
1, we clearly see cirrus at surface brightness levels of $\mu_V \sim
28$ \magsec (Janowiecki \etal in preparation).

To correct for the contamination of the images by the galactic cirrus,
we need both a high resolution, wide field map of the dust, as well as
a good model for the backscattering of the galactic light (see Witt
\etal 2008). Existing dust maps from infrared surveys suffer from poor
spatial resolution ($\sim$ 6\arcmin\ in the maps of Schlegel \etal
1998); while {\it Spitzer} provides higher spatial resolution than
IRAS ($\sim$ 1\arcmin), its small field of view makes deep imaging
over a wide area such as the Virgo cluster prohibitively expensive.
As such, the galactic cirrus represents a hard floor to deep broadband
surveys for intracluster light. To go deeper, one must rely on star
counts (\eg Ferguson \etal 1998; Ferguson \etal 2002; Durrell \etal
2002; Williams \etal 2007) which at the distance of Virgo can
currently be done only over small angular scales using space-based
telescopes. Again, mapping the ICL across the face of Virgo this way
will be prohibitive.

In summary, we find a modest correlation on large ($\sim$ 100 kpc)
scales between the broad-band diffuse light in Virgo and the inferred
density of IPNe. There is appreciable scatter to this correlation, and
the correlation depends appreciably on the underlying surface
brightness. In regions of higher surface brightness near luminous
ellipticals, the connection is evident, and we are likely seeing a PNe
population which simply traces the luminous profile of the the
connection shows appreciable scatter on large scales, and no
correlation on smaller ($\sim$ 10 kpc) scales. A number of
observational and physical effects act to wash out the expected
``IPNe-follow-light'' correlation signature, the most pernicious of
which are contamination the IPN catalogs and the small number
statistics that arise from the inherently small number of PNe per unit
luminosity in any stellar population.  Indeed, given these
complications, it is encouraging that we recover even the weak
correlations seen in our data, and our results remain consistent with
the expectation that IPNe trace the underlying broadband
ICL. Improvements on this analysis will need to come through improved
IPN samples which go deeper (to increase sample sizes) {\it and}
simultaneously reduce the contamination due to background emission
sources and faint stellar sources.

\acknowledgments

We thank Ortwin Gerhard, Magda Arnaboldi, and their collaborators for
providing their IPN catalogs.  This work is supported by NSF grants
AST-9876143 and AST-0607526 (JCM), AST-0098435 (HLM), and AST-0302030
(JJF), and by Research Corporation Cottrell Scholarships to JCM and
HLM.

\appendix
\section{The Error Model}

We quantify the photometric uncertainties in our broadband imaging,
following the prescription of Morrison \etal 1997, and direct the
reader there for more detailed descriptions of the process.

The first set of considerations are processes that introduce
uncorrelated noise on a pixel-by-pixel basis: readnoise, photon noise,
and small-scale flat fielding errors. For the SITe CCD used in the
observations, the gain was 2.8e$^-$/ADU and the readout noise was
7.5e$^-$, or 2.6 ADU. Since the ICL is a small fraction ($<$1\%) of the
night sky, photon noise is dominated by sky photons. For a typical sky
background of 1250 ADU, we expect photon noise of 21 ADU per pixel
in a single image. Photon noise in the individual blank sky images
will also translate into uncertainty in the resulting flat field. This
uncertainty will be mediated in the final image by our dithering
pattern -- each pixel in the final image comes from a different pixel
of the flat field (each with independent errors). The final error is
given by 
\begin{equation} \sigma_{sff} = \frac{\sqrt{C_{s}}}{\sqrt{g}}
\frac{1.22}{\sqrt{N_{f}}}
\end{equation}
where $C_{s}$ is the number of counts in the final, combined master
sky flat image, g is the gain, $N_{f}$ is the number of individual sky
flats used to make the master sky flat. For a typical sky value of
1250 ADU, and 127 individual sky images, this corresponds to 2.3 ADU.

These processes will sum in quadrature to give an uncertainty in a
single Virgo image of 21.4 ADU per pixel. Since the final mosaic
consists of a median of many individual images, this noise will scale
with image number as $1.22/\sqrt{N_i}$, where $N_i$ is the number of
individual images that contribute to the final mosaic. Due to our
large dithering pattern, this number varies across the mosaic. For
most of the IPN fields, $\sim$ 35 images contribute to that portion of
the mosaic (see Table 1); Field 7 has the most ($\sim$ 45 images), and
LPC the fewest ($\sim$ 17 images). For $N_i=40$, the random error per
pixel becomes 4.5 ADU, while for $N_i=17$ it becomes 6.6 ADU. When
measuring the ICL over large fields, such as the IPN survey boxes
shown in Figure 1, these errors drop as $\sqrt{N_{pix}}$ and become
negligible when compared to the uncertainty due to large scale flat
fielding and sky background variations (discussed below). However,
when measuring the ICL on small scales (as done in \S4), the
measurement boxes are smaller and the uncertainty can be
significant. The smallest boxes we analyze are 3x3 pixels in size;
this corresponds to an uncertainty of 1.4 (2.2) ADU for 40 (17)
combined images.

Over larger scales, there are two dominant sources of uncertainty
which do not easily ``root N'' away. The first is large scale flat
fielding errors, predominantly due to variation in sky brightness and
the unsubtracted extended wings of stars in the blank sky images. The
second is night sky variations in the object images themselves, which
can vary both randomly and systematically (with, for example, hour
angle, time of night, or airmass). To quantify these combined
uncertainties, we identify regions of blank sky (predominantly in the
southwest, north, and eastern portions of the image) and measure the
residual flux and uncertainty in 100x100 pixel boxes. These boxes show
a mean residual sky background of 0.1 ADU and a variation of 0.5 ADU,
which we adopt as our measure of the large scale uncertainty of our
data. Since these effects do not ``root N'' away, they represent the
floor to our error model.

\end{document}